\documentclass[%
 reprint,
superscriptaddress,
 amsmath,amssymb,
 aps,
 prl,
floatfix,
]{revtex4-2}

\usepackage{float}
\usepackage{xcolor}
\usepackage{graphicx}% Include figure files
\usepackage{dcolumn}% Align table columns on decimal point
\usepackage{bm}% bold math
\usepackage{comment}
\usepackage[unicode=true,pdfusetitle,bookmarks=false,colorlinks=true,citecolor=blue,urlcolor=blue,linkcolor=blue]{hyperref}

\begin{document}

\preprint{APS/123-QED}

\title{Kinetic temperature and pressure of an active Tonks gas}% Force line breaks with \\

\author{Elijah Schiltz-Rouse}
\affiliation{Department of Chemistry, The Pennsylvania State University, University Park, Pennsylvania, 16802, USA}

\author{Hyeongjoo Row}
\email{hrow@berkeley.edu}
\affiliation{Department of Chemical and Biomolecular Engineering, UC Berkeley, Berkeley, CA 94720, USA}

\author{Stewart A. Mallory}
\email{sam7808@psu.edu}
\affiliation{Department of Chemistry, The Pennsylvania State University, University Park, Pennsylvania, 16802, USA}


\date{\today}% It is always \today, today,

\begin{abstract}
Using computer simulation and analytical theory, we study an active analog of the well-known Tonks gas, where active Brownian particles are confined to a periodic one-dimensional (1D) channel.
By introducing the notion of a kinetic temperature, we derive an accurate analytical expression for the pressure and clarify the paradoxical behavior where active Brownian particles confined to 1D exhibit anomalous clustering but no motility-induced phase transition.
More generally, this work provides a deeper understanding of pressure in active systems as we uncover a unique link between the kinetic temperature and swim pressure valid for active Brownian particles in higher dimensions. 
\end{abstract}

\maketitle

A fundamental model for understanding the behavior of liquids is the one-dimensional (1D) hard rod fluid or Tonks gas~\cite{Tonks1936-pf}, where purely-repulsive particles are confined to move in a 1D channel.
The statistical mechanics of this system can be solved exactly, and many of its equilibrium properties are derivable in a closed analytical form~\cite{Tonks1936-pf, Sells1953-ft, Salsburg1953-fu, Koppel1963-gf, Gursey1950-oz, Rushbrooke1948-oh, santos2016concise, hansen2013theory}.  
There is a rich history in statistical physics of studying the Tonks gas and its variations to validate theories and approximation~\cite{Kac1963-gr, Helfand1961-gx, Rowlinson1965-yz, Robledo1986-ym, Andrews1975-aq, Pergamenshchik2020-aq, Lenard1961-vk, Cecconi2004-qq}. 
More recently, 1D models have achieved elevated importance as meaningful representations of physical systems, including the single-file diffusion of colloids in microfluidic channels, ion transport across cellular membranes, and the clustering of red blood cells in microcapillary flows~\cite{Lin2005-df, Liam_McWhirter2011-uy, Locatelli2016-nw, Coste2010-xa, Calbi2003-vd, Hansen-Goos2006-wr, Lutz2004-gy, Wei2000-si, Misiunas2019-pm}.

Here, we focus on an active variant of the Tonks gas where active Brownian particles are restricted to move in a narrow channel. 
Active Brownian particles (ABPs) are a popular minimal model for self-propelled particles as their collective behavior captures many of the features exhibited by active suspensions~\cite{Romanczuk2012-cc, Mallory2018-nn, Bechinger2016-db, Shaebani2020-pe, Gompper2020-zx, Zottl2023-sg, Jeggle2020-km, Bickmann2020-mu}. 
Understanding the active Tonks gas is paramount, as many of the proposed applications of microscopic active matter will operate in highly confined environments. 
Examples include targeted drug delivery to specific cellular targets~\cite{Din2016-id, Ghosh2020-ys, Patra2018-bg} and the autonomous exploration of porous media~\cite{Kumar2022-zm, Bhattacharjee2021-et}.
Also, active particles tend to accumulate on surfaces, leading to the formation of 1D boundary layers~\cite{Yan2018-fp,Jamali2018-fy,Yan2015-lk,Duzgun2018-yg,Razin2020-wk,Fily2014-do,Ostapenko2018-wo,Das2019-sp,Turci2021-sp,Ketzetzi2022-wl}.

Surprisingly, there is little work exploring the role of pressure in 1D active systems. Nevertheless, pressure is critical in characterizing the behavior of active systems~\cite{Winkler2015-qb, Mallory2014-gd, Solon2015-lt, Takatori2014-xz, Solon2015-hu, Smallenburg2015-hy, Speck2016-in, Joyeux2016-tu, Marini_Bettolo_Marconi2017-ql, Epstein2019-lz, Omar2020-kr, Marini_Bettolo_Marconi2016-ak, Klamser2018-ix},
and has aided in explaining a variety of phenomena, including motility-induced phase separation~\cite{o2021introduction, omar2022mechanical, Mallory2021-ux, Fily2012-st, Redner2013-ok, Wittkowski2014-in, Takatori2015-ki, Solon2018-om,Paliwal2018-my, Hermann2021-hq, Speck2021-ax, omar2022mechanical}, active particle motion within vesicles and droplets~\cite{Vutukuri2020-ep, Wang2019-ap, Gandikota2023-tg, Iyer2022-ci, Sokolov2018-yx, Takatori2020-rh, Xie2022-sf, Angelani2019-rq} and the dynamics of passive colloidal structures in an active bath~\cite{Liu2020-qv, Baek2018-am, Harder2014-tu, Feng2021-og, Li2022-tn, Zaeifi_Yamchi2017-uy, Ni2015-ea, Mallory2014-qn, Mallory2015-fy, Gandikota2022-he, Nikola2016-vw, Harder2014-wz, Mallory2020-cx, Szakasits2017-hq, Omar2019-xe, Kaiser2014-iz,Xia2019-ry, Szakasits2019-bx, Saud2021-iq}.
In this work, we introduce the concept of a kinetic temperature to aid in deriving an accurate analytical expression for the pressure of an active Tonks gas.
Inspiration is taken from the study of granular matter, where a granular temperature can be defined to account for the inelastic nature of collisions~\cite{Cecconi2004-qq}.
We find excellent agreement between our analytical result and numerical simulation.
In addition, we derive an exact relation between the kinetic temperature and the so-called swim pressure~\cite{Fily2012-st, Takatori2014-xz} valid for ABPs in any dimension.
To conclude, we explain the unusual clustering observed in 1D active systems and the absence of a motility-induced phase transition.
\begin{figure}[t!]
	\centering
	\includegraphics[width=.49\textwidth]{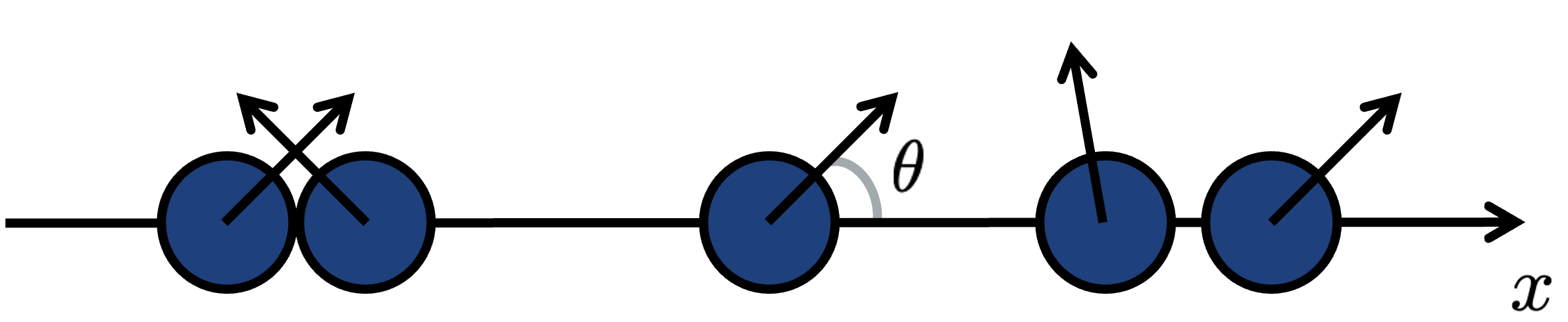}
	\caption{\protect\small{Schematic of 1D-ABP system. Each purely-repulsive particle moves at a speed $U_a \cos \theta$ while undergoing rotational Brownian motion with reorientation time $\tau_R$.}
    }
	\label{figure_1}
\end{figure}

To model the active Tonks gas or 1D-ABP system, we consider $N$ purely-repulsive active Brownian disks confined to a periodic 1D channel of length $L$ as shown in Fig.~\ref{figure_1}. 
An active force $F_{a} = \gamma U_{a}\cos{\theta}$ is applied to each particle where $\theta$ is the angle between the positive x-axis and the particle's orientation vector, $\gamma$ the translation drag coefficient, and $U_a$ the constant self-propelling speed.
All particles undergo rotational Brownian motion with a characteristic reorientation time $\tau_R$, {  where the particle's orientation vector is restricted to the two-dimensional plane parallel to the channel.
Thus}, the motion of each particle is governed by the overdamped set of equations:
\begin{subequations}
    \label{eq:1}
    \begin{equation}
        v = \dot{x} = \gamma^{-1}(F_{a} + F_{c}) = U_a \cos(\theta) + \gamma^{-1}F_c \ ,
        \label{subeq:1a}
    \end{equation}
\begin{equation}
        \dot{\theta} = \xi(t) \ ,
        \label{subeq:1b}
    \end{equation}
\end{subequations}
\noindent where $F_c$ is the interparticle force and $\xi(t)$ is Gaussian white noise characterized by $\langle \xi(t) \rangle = 0$ and ${\langle \xi(t)\xi(t') \rangle = (2/\tau_{R})\delta(t-t')}$.
We consider the limit where the effects of translational Brownian motion are nominal.
The interparticle force $F_c$ arises from a Weeks-Chandler-Anderson (WCA) potential characterized by a potential depth $\varepsilon$ and Lennard-Jones diameter $\sigma$~\cite{Weeks1971-kp}.
As our active force is bounded, a sufficiently steep potential mimics a hard-particle interaction.
A choice of $\varepsilon/(F_a \sigma) = 100$ results in hard particle statistics with a particle length of $\sigma_p = 2^{1/6} \sigma$.
Using the \texttt{HOOMD-blue} software package~\cite{Anderson2020-wl}, all simulations were conducted with 1,000 particles and run for a minimum duration of 5,000~$\sigma/U_0$. 

Two dimensionless parameters describe the state of the purely-repulsive 1D-ABP system: the packing fraction $\phi = \rho \sigma_p$ where $\rho=N/L$ is the particle line density and the dimensionless run-length $\ell_0 = (U_a \tau_R) / \sigma$. 
In the limit of small run-lengths, the behavior of the 1D-ABP system recovers that of the equilibrium Tonks gas.
The emergent behavior that arises as run-length increases is the formation of large dynamic clusters.
Cluster formation appears to be a universal feature of 1D active matter systems and has been previously observed in several studies~\cite{Locatelli2015-cs,Dolai2020-cu,De_Castro2021-fa,Gutierrez2021-mu,Sepulveda2018-gf,Soto2014-if,Sepulveda2016-sr,Slowman2016-gx,Caprini2018-xx,mukherjee2022nonexistence}.
To quantify cluster formation, we define an empirical measure of the degree of clustering $\Theta = 1 - 1/\langle N_c \rangle$, where $\langle N_c \rangle$ is the average number of particles in a cluster and $\langle ... \rangle$ denotes a time average. 
{  Here, particles are considered clustered when in contact (i.e., the separation distance is less than $\sigma_p$).}
If there are predominately unclustered particles in the system $\Theta \approx 0$, while if all particles belong to a single cluster $\Theta = 1 - 1/N \approx 1$. 

\begin{figure}[t!]
	\centering
	\includegraphics[width=.49\textwidth]{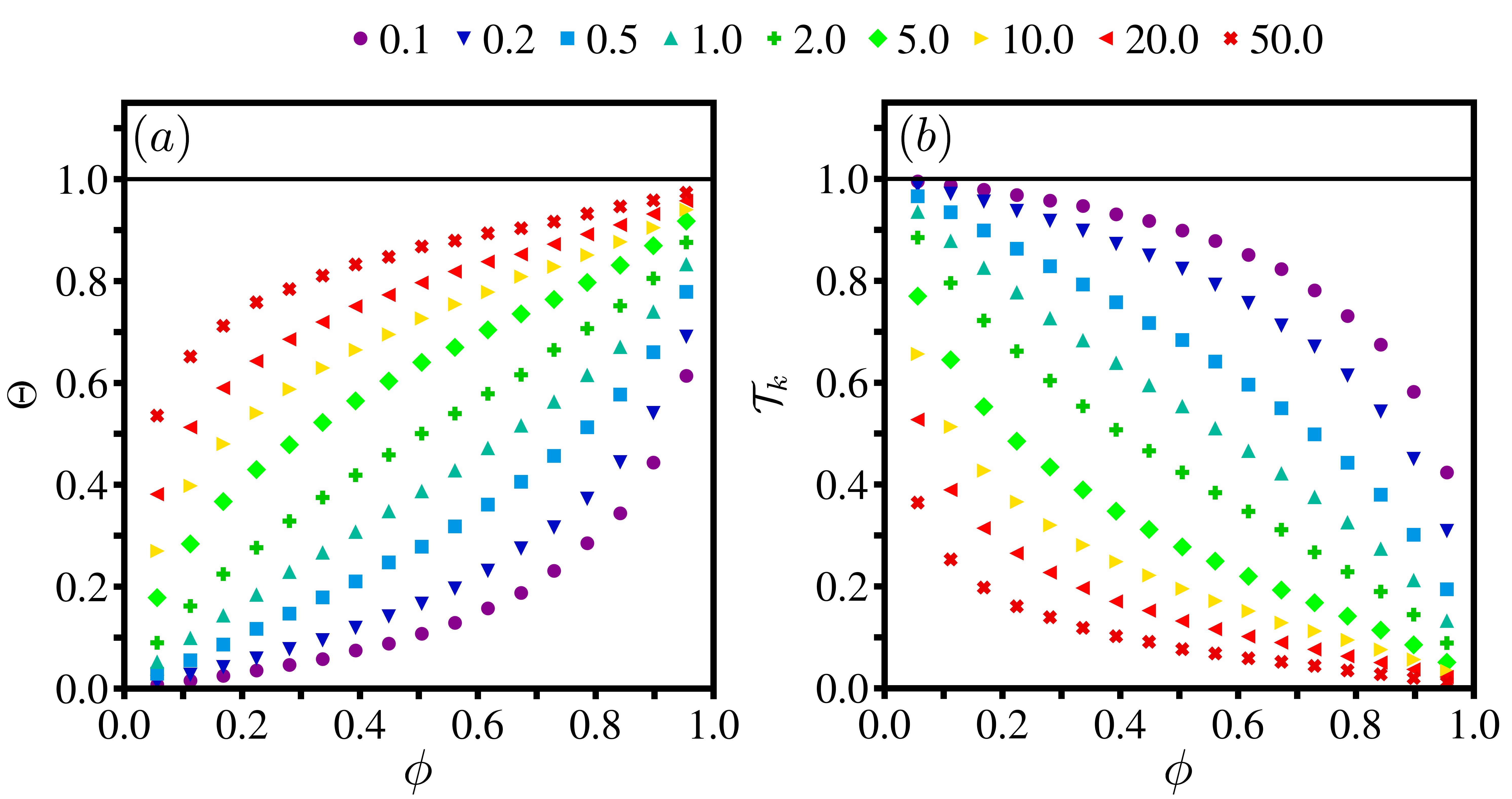}
	\caption{\protect\small{(a) Degree of clustering $\Theta$ and (b) reduced kinetic temperature $\mathcal{T}_k$ as a function of packing fraction $\phi$ for different run-lengths $\ell_0$.}}
	\label{figure_2}
\end{figure}

In Fig.~\ref{figure_2}(a), we plot the degree of clustering $\Theta$ for the 1D-ABP system as a function of packing fraction $\phi$ for different values of $\ell_0$.
The degree of clustering increases with the packing fraction and approaches $\Theta = 1$ at close-packing (i.e., $\phi =1$).  
In the opposite limit as $\phi \rightarrow 0$, there is little clustering and $\Theta \rightarrow 0$.
Yet, it is notable that $\Theta$ can increase dramatically even at low packing fractions when $\ell_0$ becomes large.
In the Supplemental Material~\footnote{See Supplemental Material at [URL] for details.}, we include movies from simulations illustrating the cluster dynamics at different run-lengths. 

A notable observation for the 1D-ABP system is that particles' velocities are significantly reduced when clustered. 
To quantify this reduction in particle speed, we define a reduced average translational kinetic energy: $\mathcal{K} = K/K_0 = 2 \langle v^2 \rangle/ U_a^2$, where $K$ and $K_0$ are the kinetic energies of the interacting and ideal 1D-ABP system, respectively.
It is worth noting the reduced average translation kinetic energy $\mathcal{K}$ is closely related to the dissipation or irreversible energy loss typically defined in stochastic thermodynamics~\cite{Sekimoto1998-cy, Seifert2012-cl}.
Interestingly, recent work has linked dissipation to the structure and transport properties of active liquids~\cite{Rassolov2022-ak,Del_Junco2018-xu,Tociu2019-mv}.

For the 1D-ABP system, it is easy to show from Eq.~(\ref{eq:1}) that $\mathcal{K} = 1 - 2 \langle F_c^2 \rangle/(\gamma U_a)^2$.
The second term represents the reduction in kinetic energy due to interparticle collisions.
In deriving $\mathcal{K}$, we use a unique property of ABPs, which we prove in the Supplemental Material~\cite{Note1}, where $\langle vF_c \rangle = 0$.
A consequence of this property is $\langle F_aF_c \rangle = -\langle F_c^2 \rangle$ for all packing fractions, run-lengths, and system sizes.
Remarkably, this property is not a result of 1D confinement, but the analog $\langle \bm{v} \cdot \bm{F}_c \rangle = 0$ is also true for ABPs in higher dimensions.

For a 1D system in thermal equilibrium, the temperature $T$ can be obtained by application of the equipartition theorem to give the well-known result $k_BT = 2 K/N$, where $k_B$ is the Boltzmann constant. 
In a similar spirit, we define a kinetic temperature for ABPs as $T_k = 2 K/N$, resulting in a reduced kinetic temperature:
\begin{equation}
\label{eq:2}
\mathcal{T}_k = \frac{T_k}{T_0} = 1 - \frac{2 \langle F_c^2 \rangle}{(\gamma U_a)^2}\ , 
\end{equation}
where $T_{0}$ is the kinetic temperature of the ideal active system, and the second term is the relative reduction of the kinetic temperature due to collisions.
\begin{figure*}[t!]
%	\centering
	\includegraphics[width=0.975\textwidth]{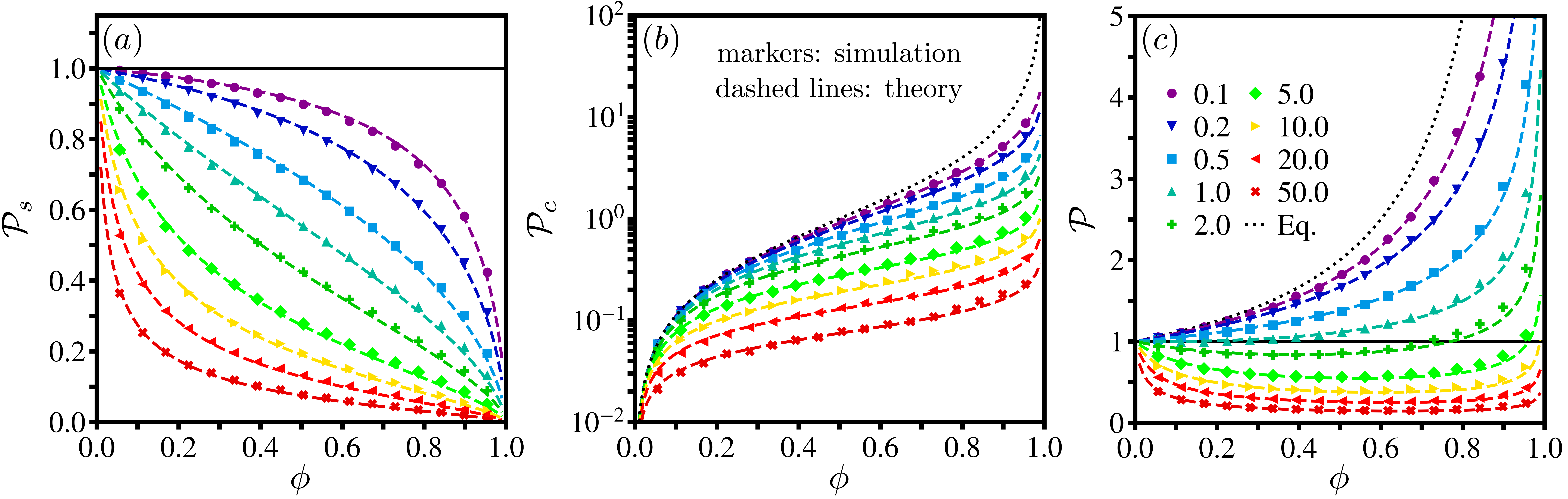}
	\caption{\protect\small{Brownian dynamics simulation results for the reduced (a) swim pressure, (b) collisional pressure, and (c) total pressure as a function of packing fraction for different run-lengths. The dotted lines correspond to the analytical expressions for the Tonks gas. The dashed curves are the analytical expression derived based on the scaling argument for the kinetic temperature.} }
	\label{figure_3}
\end{figure*}

In Fig.~\ref{figure_2}(b), we plot $\mathcal{T}_k$ for the 1D-ABP system. 
The reduced kinetic temperature decreases monotonically as the packing fraction increases and, in agreement with our previous observation, scales inversely with the degree of clustering shown in Fig~\ref{figure_1}(a).
At low packing fraction, $\mathcal{T}_k$ approaches the ideal result, and at close-packing, $\mathcal{T}_k$ tends to zero for all run-lengths.
This temperature reduction or kinetic energy loss is a peculiarity of active Brownian systems.
The inelastic nature of collisions reduces a particle's velocity during a collision and, in turn, temporarily removes kinetic energy from the system. 

The utility of $\mathcal{T}_k$ is that it can be related directly to the mechanical pressure.
The total pressure for the 1D-ABP system is computed via the virial theorem as $P = \rho \langle x F_{net} \rangle$ where $F_{net} = F_a + F_c$ is the net force acting on a particle.
Here, $P$ is the average force the ABPs exert on the boundary.
The two contributions to the total pressure $P$ are the collisional pressure $P_c = \rho \langle x F_c  \rangle$ and the swim pressure $P_s = \rho \langle x F_a \rangle$.
For convenience, we use an alternative yet equivalent  expression for the swim pressure valid for ABPs known as the ``active impulse'' form given by $P_s = \rho  \langle v F_a \rangle \tau_R$~\cite{Patch2018-aa,Fily2017-uo,Das2019-cd}.  
The pressure of the ideal 1D-ABP system is $P_0= \rho \gamma U_a^2\tau_R / 2$, and the reduced swim pressure of the interacting system is
\begin{equation}
\label{eq:3}
\mathcal{P}_s = \frac{P_s}{P_0} = 1 - \frac{2\langle F_c^2 \rangle}{(\gamma U_a)^2}  \ ,
\end{equation}
where we again use the relation~$\langle F_aF_c \rangle = -\langle F_c^2 \rangle$. 
Upon comparison of Eq.~(\ref{eq:2}) and Eq.~(\ref{eq:3}), we find the two expressions are identical and arrive at one of the central results of this work, which is the equivalence of the reduced swim pressure and reduced kinetic temperature: $\mathcal{P}_s = \mathcal{T}_k$.
This equivalence can be easily extended to ABPs in higher dimensions, as shown in the Supplemental Material~\cite{Note1}. 
In Fig.~\ref{figure_3}(a), we plot $\mathcal{P}_s$ computed from simulation and validate $\mathcal{P}_s = \mathcal{T}_k$ by comparison with Fig.~\ref{figure_2}(b).
A practical outcome of $\mathcal{P}_s = \mathcal{T}_k$ is that the kinetic temperature offers a convenient method of computing the swim pressure of ABPs as particle velocities are readily available from simulation.

We derive an analytical expression for the reduced kinetic temperature using a simple scaling argument predicated on the dynamics of an individual particle.
There are two relevant timescales for a given particle: the average duration of a collision $\tau_C$ and the average time between collisions $\tau_F$.
The reduced kinetic temperature can be estimated as $\mathcal{T}_k = [(1) \tau_F + (0) \tau_C]/(\tau_F+\tau_C) = 1/(1+\tau_C/\tau_F)$, where between collisions $\mathcal{T}_k = 1$ as particle motion is unimpeded and during a collision $\mathcal{T}_k = 0$.
We approximate the duration of a collision as being comparable to the reorientation time of a particle: $\tau_C \sim \tau_R$.
While the time between collision scales as $\tau_F \sim \lambda/U_a$, where $\lambda$ is the mean free path and $U_a$ is the intrinsic speed of a particle. 
The mean free path in the limit of small run-length is $\lambda = (1-\phi)/\rho$ as particles are nearly uniformly distributed. For the more general case, $\lambda \sim (1-\phi)/(\rho\sqrt{\mathcal{T}_k})$ as clustering increases the distance a particle must travel between collisions (see Supplemental Material~\cite{Note1}). A full discussion of the dependence of $\tau_C$ and $\tau_F$ will be published separately.

This expression captures the asymptotic behavior of the reduced kinetic temperature in Fig.~\ref{figure_2}(b) (i.e., $\mathcal{T}_k  \rightarrow 1$ as $\tau_C/\tau_F \rightarrow 0$ and $\mathcal{T}_k  \rightarrow 0$ as $\tau_C/\tau_F \rightarrow \infty$).
We expect the exact value of the kinetic temperature will be sensitive to the assumption that particle velocities are exactly zero when clustered.
We recognize this is not strictly true as we observe that clusters retain a small drift velocity. 
To account for this drift behavior and the approximation of $\tau_C$ and $\tau_F$, we introduce a correction factor $\alpha$, to be determined \textit{a posteriori}, and obtain an algebraic equation for the reduced kinetic temperature:
\begin{equation}
\label{eq:4}
\mathcal{T}_k = \left(1+\alpha \frac{\tau_C}{\tau_F} \right)^{\!-1} = \left(1+\alpha \ell_0 \frac{{\color{black}\phi}}{1-\phi}\sqrt{\mathcal{T}_k} \right)^{\!-1}  \ .
\end{equation}
By solving Eq.~(\ref{eq:4}) (see Supplemental Material~\cite{Note1}), we obtain the following analytical expression for $\mathcal{T}_k$:
\begin{equation}
\mathcal{T}_k = \frac{1}{9b^2}\left[2 \cos \left(\frac{1}{3}\arccos\left(\frac{27}{2} b^2 - 1\right)\right) - 1\right]^2 \ ,
\label{eq:5}
\end{equation}
where $b = \alpha \ell_0 \phi /(1-\phi)$.
In Fig.~\ref{figure_3}(a), we find excellent agreement between simulation results and our analytical solution for $\mathcal{T}_k$ or equivalently $\mathcal{P}_s$. 
As expected, $\alpha$ exhibits a weak dependence on $\ell_0$ and can be approximated by $\alpha = c/(1+\ell_0)^d$ where $c = 1.1$ and $d=0.05$.
 
We now derive an analytical expression for the collisional pressure of the 1D-ABP system.
As the run-length decreases, the collisional pressure [Fig.~\ref{figure_3}(b)] approaches the known analytical result for the equilibrium system: $\mathcal{P}_c = P_c/P_0 = \phi/(1-\phi)$.
For particles interacting through an additive pairwise potential, the collisional pressure can be expressed as $P_c = \rho \langle x_{ij} F_{ij} \rangle/2$ where $x_{ij} = x_j - x_i$ is the distance between the $i^{th}$ and $j^{th}$ particles and $F_{ij}$ is the resulting force between the pair.
As we consider nearly hard particle interactions, a suggestive scaling for the collisional pressure is $P_c =\rho \langle x_{ij} F_{ij} \rangle /2 \sim \rho (\sigma_p \overline{F}_{ij}) /2 $, where $\overline{F}_{ij}$ is the average magnitude of the force experienced between a pair of particle.
To recover the equilibrium result in the limit of small run-lengths, it is required $\overline{F}_{ij} = \gamma U_a^2 \tau_R [\rho/(1-\phi)] = \mathcal{F}_a  \tau_C/\tau_F$, where $\mathcal{F}_a = \gamma U_a$.
This expression for $\overline{F}_{ij}$ can be generalized to large run-lengths by replacing the force scale $\mathcal{F}_a$ by a more appropriate force scale $\mathcal{F}_a = \gamma U_a\sqrt{\mathcal{T}_k}$, which captures the observed reduction in $\mathcal{P}_c$ as $\ell_0$ is increased [see trend in Fig.~\ref{figure_3}(b)].
Thus, an expression for $P_c$ valid for all run-lengths is
\begin{equation}
  P_c =\! \frac{\rho}{2} \left(\sigma_p \gamma U_a\sqrt{\mathcal{T}_k} \frac{\tau_C}{\tau_F} \right) = P_s \left[ \frac{\phi}{1-\phi} \right] \, .
  \label{eq:6}
\end{equation}

\begin{figure}[t!]
	\centering
	\includegraphics[width=.49\textwidth]{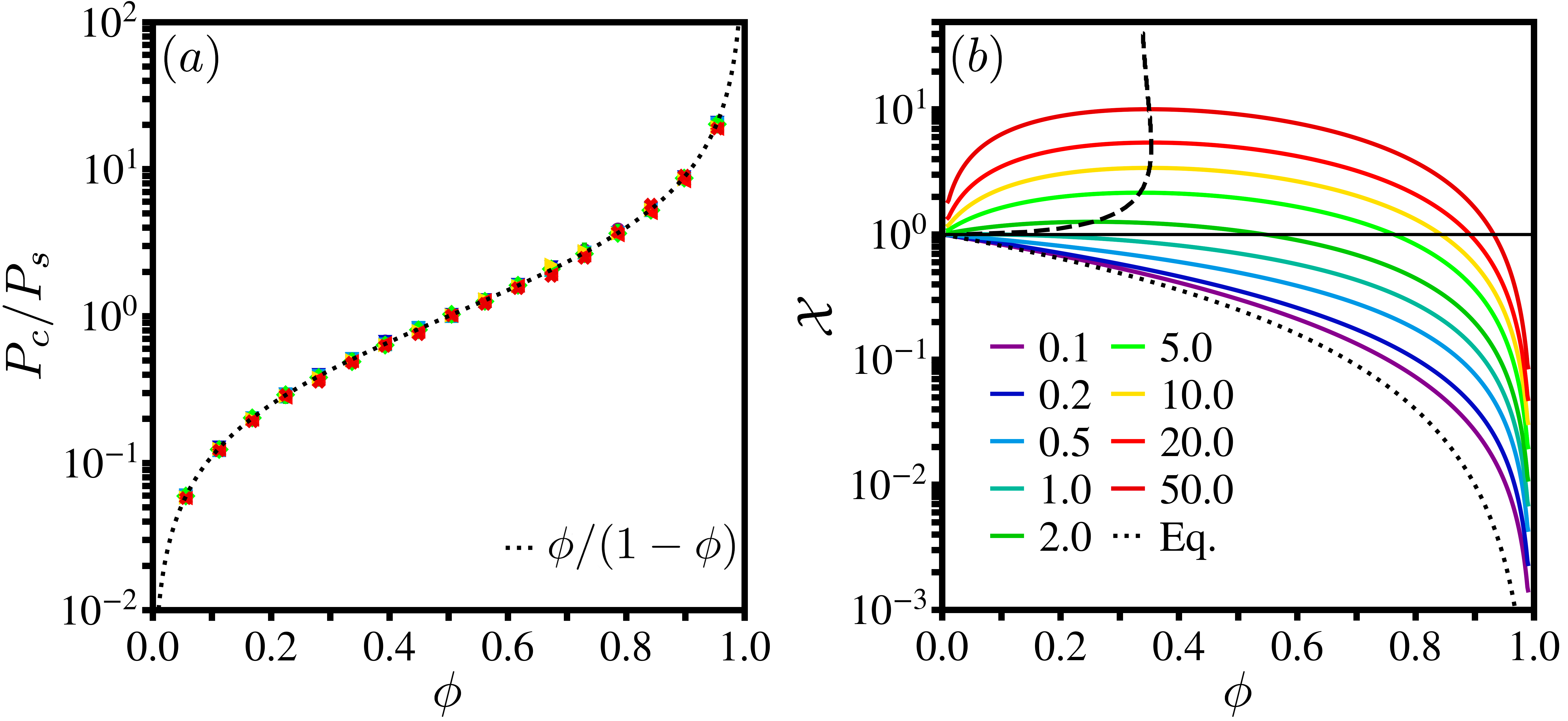}
	\caption{\protect\small{The ratio of the collisional pressure $P_c$ to the swim pressure $P_s$ for the 1D-ABP system collapse onto a universal curve $P_c/P_s = \phi/(1-\phi)$. (b) The compressibility as a function of packing fraction for various run-lengths. The dotted line corresponds to the reduced compressibility of the equilibrium Tonks gas $\mathcal{X} = (1-\phi)^2$. The dashed line traces out the location of the maximum of the reduced compressibility $\mathcal{X}_m$.}}
	\label{figure_4}
\end{figure}

In Fig.~\ref{figure_3}(b), we see excellent agreement between simulation results and the analytical expression: ${\mathcal{P}_c\!=\!\mathcal{T}_k [\phi/(1-\phi)]}$.
Remarkably, the reduced collisional pressure is simply the product of $\mathcal{T}_k$ and the equilibrium collisional pressure of the Tonks gas.
A unique feature of the 1D-ABP system, highlighted in Fig.~\ref{figure_4}(a), is the ratio of the collisional and swim pressure collapses onto a universal curve given by $P_c/P_s = \phi/(1-\phi)$.
This behavior is a direct result of the single-file confinement and not observed for ABPs in higher dimensions~\cite{Omar2021-gn,Arnoulx_de_Pirey2019-me,Dey2022-wt,Digregorio2018-mx, Levis2017-cu}.

By combining our results for the swim and collisional pressure, we achieve our primary aim of an analytical expression for the total pressure of the 1D-ABP system:
\begin{equation}
\label{eq:7}
\mathcal{P} =\mathcal{P}_s + \mathcal{P}_c = \mathcal{T}_k\left[1 + \frac{\phi}{1-\phi}\, \right] = \mathcal{T}_k\left[\frac{1}{1-\phi}\, \right] \ .    
\end{equation}
In Fig.~\ref{figure_3}(c), we find excellent agreement between our analytical expression for $\mathcal{P}$ and simulation results.
In the limit of small run-lengths, Eq.~(\ref{eq:7}) reproduces the  analytical result for the equilibrium Tonks gas: $\mathcal{P} = 1/(1-\phi)$.
We also identify an analog to the Boyle temperature of an equilibrium system, which we call the Boyle run-length, where the second virial coefficient is zero and $\mathcal{P} \approx 1$.
The Boyle run-length occurs at $\ell_0 \approx 1$ and indicates a crossover from a regime where the interaction between particles are predominantly repulsive to one where there is an effective attraction between particles. 
This crossover is consistent with the activity-induced clustering shown in Fig.~\ref{figure_2}(a). 
{  It is straightforward to show from Eq.~(\ref{eq:7}) that the single homogeneous phase for the 1D-ABP system is always mechanically stable (i.e. $(\partial P/ \partial \rho)_{\ell_0} > 0$).}
These findings provide a mechanical interpretation to prior studies on 1D active systems where there was found to be clustering but no motility-induced phase separation~\cite{Locatelli2015-cs,Dolai2020-cu,De_Castro2021-fa,Gutierrez2021-mu,Sepulveda2018-gf,Soto2014-if,Sepulveda2016-sr,Slowman2016-gx,Caprini2018-xx,mukherjee2022nonexistence}.

We can further quantify the clustering behavior by investigating the constant run-length compressibility -- a thermodynamic-like response function that provides a measure of clustering and local density fluctuations~\cite{Dulaney2021-pg,Chakraborti2016-tm}. 
In Fig.~\ref{figure_4}(b), the reduced constant run-length compressibility was calculated from Eq.~(\ref{eq:7}) as
\begin{equation}
\mathcal{X} \! = \! \frac{\chi_{a}}{\chi_{0}} \!= \! \left[\frac{\mathcal{T}_k}{(1-\phi)^2}+\frac{\phi}{1 - \phi} \left(\frac{\partial \mathcal{T}_k}{\partial \phi} \right) \right]^{-1} \ ,
\label{eq:8}
\end{equation} 
where $\chi_a =  (\partial \ln \rho / \partial P)_{\ell_0} $ and $\chi_{0} = 1/P_0$ are the compressibility for the interacting and ideal system, respectively.
In the limit of small run-lengths, the 1D-ABP system approaches the result of the equilibrium Tonks gas $\mathcal{X} = (1-\phi)^2$ and in the ideal limit where $\phi \rightarrow 0$, $\mathcal{X} \rightarrow 1$.
Above the Boyle run-length, $\mathcal{X}$ is no longer monotonic but exhibits a maximum $\mathcal{X}_{m}$.
The existence of this maximum suggests a structural transition or weak thermodynamic singularity similar to the Frenkel or Widom line of supercritical fluids~\cite{Bolmatov2013-ab, McMillan2010-er,Simeoni2010-lz, Xu2005-nr} and will be characterized in future work. 

In the limit of large run-length, the location of $\mathcal{X}_{m}$ becomes independent of $\ell_0$ and analytically can be shown to asymptotically approach a packing fraction $\phi = 1/3$ as shown by the dashed line in Fig.~\ref{figure_4}(b).
For these large values of the run-length, $\mathcal{X}_{m}$ exhibits a power law dependence given by $\mathcal{X}_{m} \approx 0.9 (\ell_0)^{19/30}$.
If we consider $\mathcal{X}/\mathcal{X}_{m}$ for these large run-lengths (See Supplemental Material~\cite{Note1}), we observe a collapse onto a universal curve similar in shape to that of $\ell_0=50$ in Fig.~\ref{figure_4}(b).
The 1D-ABP system has the interesting property that the compressibility can be made arbitrarily large by increasing the run-length, but there is no emergent singularity consistent with a critical point as the shape of $\mathcal{X}$ stops evolving in the limit of large run-lengths.
A physical interpretation of this behavior is that density fluctuations and clusters can become arbitrarily large, but it is impossible to form a large stable cluster that would give rise to a new dense phase. 

\textit{Conclusions.}-- This work derives an accurate analytical expression for the mechanical pressure of a purely-repulsive 1D-ABP system using the concept of kinetic temperature. 
By analyzing trends in the pressure, we obtain a mechanical interpretation for the phase behavior of the 1D-ABP system and the lack of a motility-induced phase transition. 
Further investigation is warranted to establish when the concept of a kinetic temperature can be extended to other active particle models, including those with hydrodynamic or electrostatic interactions.

\begin{acknowledgments}
\emph{Acknowledgments.--}
We thank Parvin Bayati, Yunhee Choi, Akin Akintunde, and Will Noid for helpful discussions and critically reading an early draft of the manuscript.
\end{acknowledgments}

%\nocite{*}

%\bibliography{references}% Produces the bibliography via BibTeX.


\end{document}

\preprint{APS/123-QED}

\title{Supplemental Material: Kinetic temperature and pressure of an active Tonks gas}% Force line breaks with \\

\author{Elijah Schiltz-Rouse}
\affiliation{Department of Chemistry, The Pennsylvania State University, University Park, Pennyslvania, 16802, USA}

\author{Hyeongjoo Row}
\email{hrow@berkeley.edu}
\affiliation{Department of Chemical and Biomolecular Engineering, UC Berkeley, Berkeley, CA 94720, USA}

\author{Stewart A. Mallory}
\email{sam7808@psu.edu}
\affiliation{Department of Chemistry, The Pennsylvania State University, University Park, Pennyslvania, 16802, USA}


\date{\today}% It is always \today, today,

\maketitle

\section{Proof of $\langle v F_c \rangle = 0$ for Active Brownian Particles}

We first present a simple cluster decomposition scheme to prove $\langle v F_c \rangle = 0$ for the one-dimensional active Brownian particle (ABP) system. 
The static correlation function of the velocity and collisional force at steady-state is defined as $\langle v F_c \rangle =\langle \sum_{i = 1}^{N} v_i F_{c,i} \rangle$ / N, where $v_i$ and $F_{c,i}$ are the velocity and collisional force of the $i^{th}$ particle, respectively, and $N$ is the total number of particles.
We rewrite the summation over individual particles as a sum over clusters:
\begin{equation} \label{eq:S1}
\langle v F_c \rangle = \frac{1}{N} \left\langle \sum_{i = 1}^{N_{\rm cl}} \sum_{j \in \mathcal{C}_i} v_j F_{c,j}\right\rangle \ ,
\end{equation}
where $N_{\rm cl}$ is the number of clusters and $\mathcal{C}_i$ is the set of indices of particles that form the $i$th cluster. 
Here, particles are considered to be clustered when the separation distance is less than $\sigma_p$, where $\sigma_p = 2^{1/6} \sigma$ is the effective hard particle length, and $\sigma$ is the standard Lennard-Jones diameter.
Due to the inelastic nature of collisions, all particles within a cluster move at the same cluster velocity $v^{\rm cl}$. 
Since particle interactions are short-ranged and conservative, the sum of interparticle forces within any cluster vanishes. 
Consequently, we obtain
\begin{equation} \label{eq:S2}
\langle v F_c \rangle
= \frac{1}{N}\left\langle \sum_{i = 1}^{N_{\rm cl}} v^{\rm cl}_{i} \sum_{j \in \mathcal{C}_i}  F_{c,j}\right\rangle = 0\ .
\end{equation}
We highlight that Eq.~\eqref{eq:S2} combined with the equation of motion $ v = \left(F_a + F_{c}\right) / \gamma$ results in $\langle F_a F_c \rangle = -\langle F_c^2 \rangle$, which establishes a bridge between the reduced kinetic temperature and reduced swim pressure as shown in the following section. 
Interestingly, this property is not a result of 1D confinement, but the analog $\langle \bm{v} \cdot \bm{F}_c \rangle = 0$ is also true for ABPs in higher dimensions. 
To prove this, we start from the fact that the rate of work performed by the active force is balanced by the rate of dissipation at steady-state: $ \langle \bm{F}_a \cdot \bm{v} \rangle = \langle \gamma \bm{v} \cdot \bm{v} \rangle$ (see Ref.~\cite{Omar2023-nn} for derivation).
By substituting the velocity given by the equation of motion $ \bm{v} = \left(\bm{F}_a + \bm{F}_{c}\right)  / \gamma$ into the prior expression, we obtain $\langle \bm{v}  \cdot \bm{F}_{c}  \rangle = 0$.

\begin{comment}
\section{Kinetic, potential, and total energy for 1D-ABP system}

\begin{figure}[H]
%	\centering
	\includegraphics[width=1.0\textwidth]{figure_S1.png}
	\caption{\protect\small{{\color{blue} (rescale)} Brownian dynamics simulation results for the (a) $\mathcal{K}$ kinetic energy per particle, (b) potential energy per particle, and (c) total energy per particle as a function of volume fraction for different run-lengths.} }
	\label{figure_S1}
\end{figure}

{\color{blue} Should we discuss this figure?} 
\end{comment}

\section{Proof of $\mathcal{T}_k = \mathcal{P}_s$ for ABPs in higher dimensions}

In the main text, we prove the equivalence of $\mathcal{T}_k = \mathcal{P}_s$ for ABPs in a single spatial dimension where particles have a single rotational degree of freedom.
For a system with higher spatial dimensions $n_S \in \{2,\ 3\}$ containing active Brownian particles with $n_R$ rotational degrees of freedom ($ n_S - 1 \leq n_R \leq 2$), the swim pressure can be written as $P_s = \rho \langle \bm{v} \cdot \bm{F}_a \rangle \tau_R / (n_S n_R)$.
It follows the swim pressure for the ideal system is $P_0 = \rho \langle \bm{F}_a\cdot \bm{F}_a \rangle  \tau_R / (n_S n_R \gamma )$ and the reduced swim pressure is
\begin{equation}
    \mathcal{P}_s = \dfrac{P_s}{P_0} = 1 - \frac{n_R + 1}{n_S}\dfrac{\langle \bm{F_c} \cdot \bm{F_c} \rangle}{(\gamma U_a)^2} \ ,
\end{equation}
where we have used $\langle \bm{F_a} \cdot \bm{F_c} \rangle = -\langle \bm{F_c} \cdot \bm{F_c} \rangle$ and $\langle \bm{F}_a\cdot \bm{F}_a \rangle = n_S \gamma^2 U_a^2 / (n_R + 1)$. The average translational kinetic energy per particle is $ K = m\langle \bm{v} \cdot \bm{v} \rangle / 2$ and for the ideal system $K_0 = m{\langle \bm{F}_a\cdot \bm{F}_a \rangle}  / (2\gamma^2) =m n_S U_a^2 / (2(n_R + 1))$. Thus the reduced kinetic energy is 
\begin{equation}
    \mathcal{K} = \dfrac{K}{K_0} = 1 - \frac{n_R + 1}{n_S}\dfrac{\langle \bm{F_c} \cdot \bm{F_c} \rangle}{(\gamma U_a)^2} \  .
\end{equation}
The kinetic temperature in $n_S$ spatial dimensions is given by $T_k = 2 K /n_s$, and it is straightforward to see $\mathcal{T}_k = \mathcal{K}$.
Thus, $\mathcal{T}_k = \mathcal{P}_s$ for the more general case of ABPs in two and three dimensions.

\section{Estimation of the run-length dependence of the mean free path}
\begin{figure}[H]
	\centering
	\includegraphics[width=0.95\textwidth]{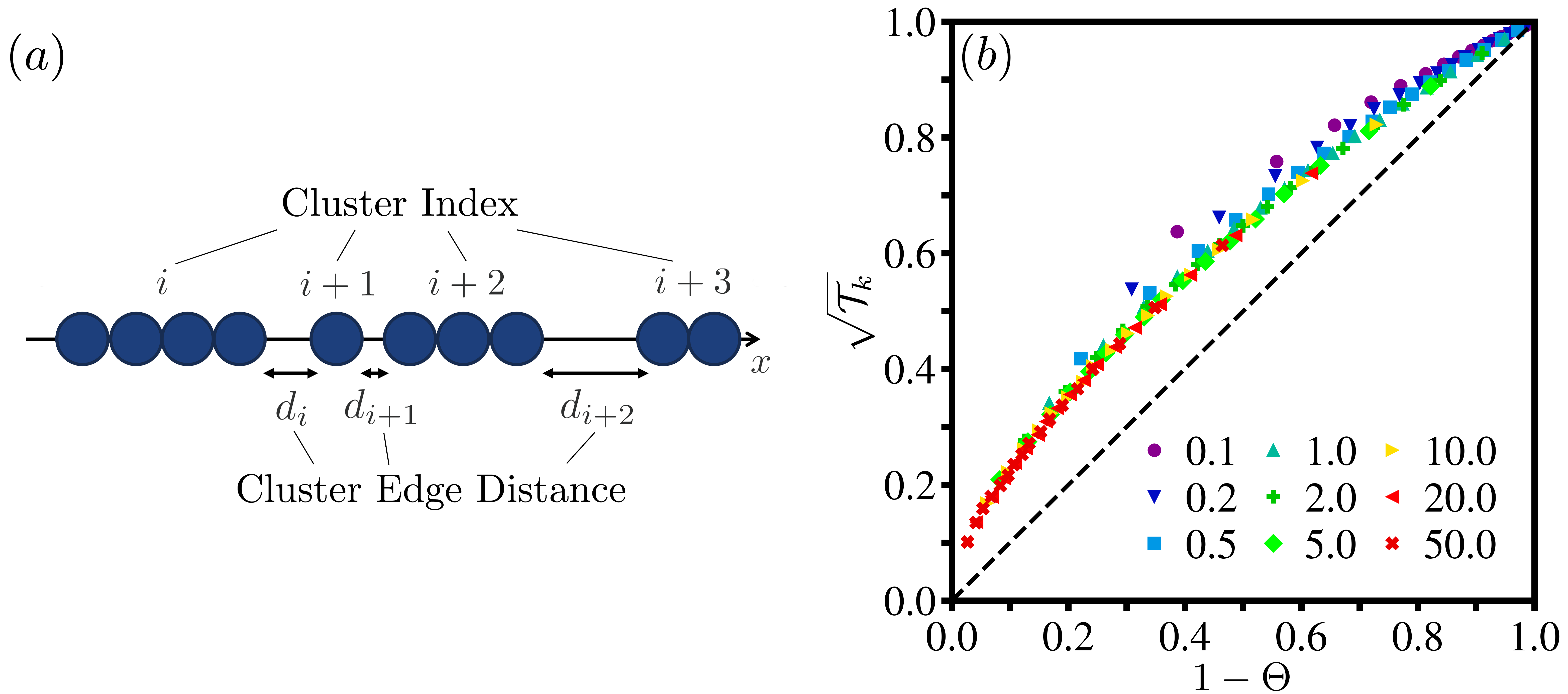}
	\caption{\protect\small{(a) Schematic of 1D-ABP system illustrating the edge distance between clusters used to approximate the mean free path. Our scheme for enumerating clusters and the corresponding edge distances are shown above. (b) The relationship between $(1- \Theta)$ and $\sqrt{\mathcal{T}_k}$} exhibit an approximately linear dependence, which was used in the final step of deriving the scaling for the mean free path [Eq.~(\ref{eq:S8})]. }
	\label{figure_Schematic}
\end{figure}

Our derivation of the reduced kinetic temperature $\mathcal{T}_k$ (Eq.~(4) of the main text) requires an estimate of the time between collision for an individual particle $\tau_{F}$. 
We can express the time between collisions as $\tau_F = \lambda/U_a$ where $U_a$ is the intrinsic speed of an ABP and $\lambda$ is the mean free path.
We estimate the mean free path as the average distance between the edges of any two adjacent clusters $\langle d \rangle$:
\begin{equation}
\label{eq:def_inter_cluster_dist}
\lambda = \langle d \rangle = \left\langle \frac{1}{N_{\rm cl}} \sum_{i=1}^{N_{\rm cl}} d_i \right\rangle \ ,
\end{equation}
where $d_i$ is the edge distance between the $i^{th}$ and $(i+1)^{th}$ cluster (See Fig.~\ref{figure_Schematic}(a)) and $N_{\rm cl}$ is the total number of clusters in a given configuration. 
As the system is periodic, we define $d_{N_{\rm cl}}$ as the distance between the farthest right cluster and the first positive periodic image of the first cluster.
The sum of all cluster edge distances is the total length that is not occupied by the particles (i.e., ${\sum d_i\!=\!L - N \sigma_p}$, where $L$ is the length of the system, $N$ is the number of particles, and $\sigma_p$ is the diameter of a particle). 
Consequently, Eq.~\eqref{eq:def_inter_cluster_dist} can be rewritten as
\begin{equation}
\label{eq:def_inter_cluster_dist2}
\lambda = \left\langle N_c \right\rangle  \frac{1 - \phi}{\rho} \ , 
\end{equation}
where $\rho=N/L$ is the particle line density, $\phi=\rho \sigma_p$ is the packing fraction, and $\langle N_c \rangle$ is the average number of particles in a cluster.
It follows from our definition for the degree of clustering $\Theta = 1 - 1/\langle N_c \rangle$, we can express the mean free path as
\begin{equation}
\label{eq:def_inter_cluster_dist3}
\lambda = \frac{1}{1 - \Theta}\frac{1 - \phi}{\rho} \ .
\end{equation}
As noted in the main text, there exists an inverse relationship between reduced kinetic temperature $\mathcal{T}_k$ and the degree of clustering $\Theta$ (See Fig. 2 of main text). 
It follows that $(1-\Theta) \sim \mathcal{T}_k^\beta$ where $\beta$ is a positive value.
In Fig.~\ref{figure_Schematic}(b), we plot  $1 - \Theta$ versus $\sqrt{T}_k$ and demonstrate an approximate linear scaling between the two quantities. 
Thus, an approximate scaling for the mean free path that captures the effects of clustering and increases with increasing run-length is 
\begin{equation}
\label{eq:S8}
\lambda \sim \frac{1}{\sqrt{T_k}}\frac{1 - \phi}{\rho} \ .
\end{equation}

\section{Analytical Expression for Kinetic Temperature}
Equation~(4) of the main text gives the algebraic equation for the kinetic temperature:
\begin{equation}
\mathcal{T}_k = \dfrac{1}{1+\alpha \ell_0 \dfrac{\phi}{1-\phi}\sqrt{\mathcal{T}_k}} = \dfrac{1}{1+b\sqrt{\mathcal{T}_k}} \  ,
\label{eq:S6}
\end{equation}
where $b = \alpha  \ell_0 \phi/ (1-\phi)$, is introduced to write Eq.~(\ref{eq:S6})  more compactly. A solution to Eq.~(\ref{eq:S6}) can be obtained by using a substitution of variables $x = \sqrt{\mathcal{T}_k}$ to transform Eq.~(\ref{eq:S6}) into a cubic equation: $bx^3+x^2-1 = 0$. 
It is straightforward to rewrite this cubic expression as a depressed cubic and use Vi\`{e}te's formula to obtain:
\begin{equation}\label{eq:eos}
\mathcal{T}_k = \frac{1}{9b^2}
\left[
                2 \cos \left(
                    \frac{1}{3}\arccos\left(\frac{27}{2} b^2 - 1\right)
                \right) - 1
\right]^2 \ .
\end{equation}

\section{Large run-length behavior of the maximum constant-run-length-compressibility}

\begin{figure}[H]
	\centering
	\includegraphics[width=0.75\textwidth]{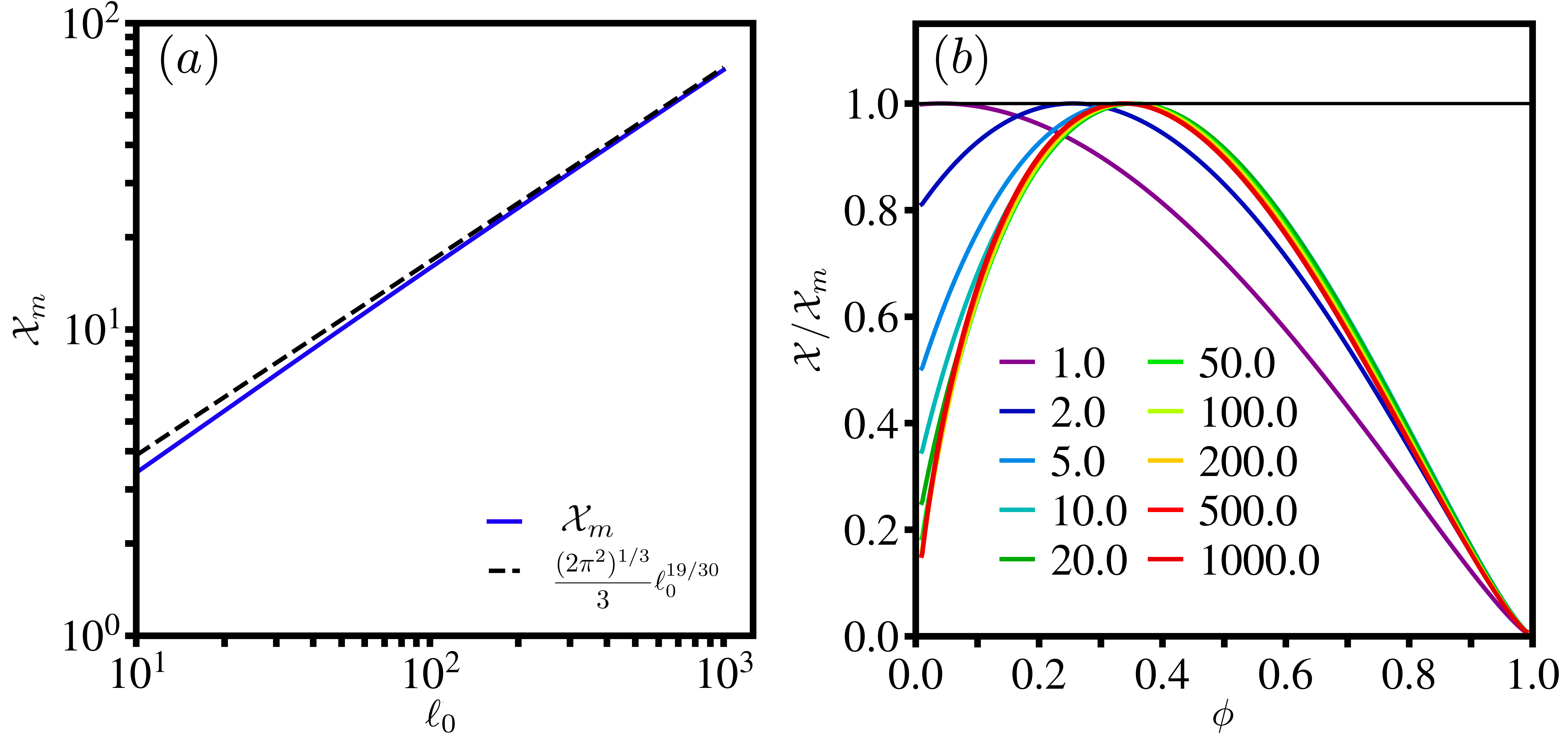}
	\caption{\protect\small{ (a) The run-length dependence of the maximum in the reduced constant-run-length compressibility $\mathcal{X}_m$, which scales as $\mathcal{X}_{m} = a (\ell_0)^{19/30}$ with $a =  (2 \pi^2)^{1/3}/3$ for large run-lengths. (b) The reduced constant run-length compressibility $\mathcal{X}$} rescaled by the maximum $\mathcal{X}_{m}$, wherein the limit of large run-length the profile of $\mathcal{X}$/$\mathcal{X}_{m}$ takes on a constant profile shape. }
	\label{figure_S2}
\end{figure}

In the main text we consider the reduced constant run-length compressibility $\mathcal{X}$ (see Eq.~(8) in the main text) to quantify the clustering behavior and present that $\mathcal{X}$ admits a maximum value $\mathcal{X}_m$ at sufficiently high run-lengths. Here, we perform a perturbative analysis on $\mathcal{X}_m$ and present its asymptotic scaling behavior at high run-lengths $\ell_0 \gg 1$. From the equation of state~\eqref{eq:eos} with a parameter $\alpha = \pi(1 + \ell_0)^{-0.05} /2^{3/2} $ computed by simulation results, in the limit of $\ell_0\rightarrow \infty$, $\mathcal{X}$ is maximized when
\begin{equation}\label{eq:S9}
3 \pi^{\frac{2}{3}} \phi^{\frac{8}{3}}\left(\phi - \frac{1}{3}\right) + 4\phi^2\left(\phi - \frac{2}{3}\right)\left(\phi - 1\right)^{\frac{2}{3}} \ell_0^{-\frac{19}{30}} + \mathcal{O}\left(  \ell_0^{-\frac{19}{15}}
\right)= 0 \ .
\end{equation}
Thus, the density $\phi_m$ at which a maximum value $\mathcal{X}_m$ is attained approaches $1/3$ as activity increases. To determine the leading order behavior in the limit of $\ell_0\rightarrow \infty$, we pose a perturbative expansion $\phi_m = 1/3 + c \ell_0^{-19/30} + \mathcal{O}( \ell_0^{-19/15} )$. From the maximum condition~\eqref{eq:S9} with the expansion, we obtain the leading order coefficient $c=4 (2/\pi)^{2/3} / 9$. By considering $\mathcal{X}$ at $\phi_m$ we find the leading order behavior $\mathcal{X}_m = (2 \pi^2)^{1/3} \ell^{19/30}/3 + \mathcal{O}(1)$ as $\ell_0\rightarrow \infty$.

\clearpage

\section{Details of Simulation Videos}

Videos of typical trajectories illustrating the clustering dynamics for different values of the volume fraction $\phi$ and run-length $\ell_0$.  
Each particle has a vector representing its orientation and the coloring is based on cluster size. 
Note that only a portion of the system is visible in each video as it would be difficult to see the entire system otherwise.
Videos are available at : 10.6084/m9.figshare.24461047

\begin{enumerate}
    \item \texttt{video\_1.mp4} -- $\phi=0.28$ and $\ell_0= 0.1$
    \vspace{-0.2cm}
    \item \texttt{video\_2.mp4} -- $\phi=0.28$ and $\ell_0= 1.0$
    \vspace{-0.2cm}
    \item \texttt{video\_3.mp4} -- $\phi=0.28$ and $\ell_0= 10.0$
    \vspace{-0.2cm}    
    \item \texttt{video\_4.mp4} -- $\phi=0.56$ and $\ell_0= 0.1$
    \vspace{-0.2cm}    
    \item \texttt{video\_5.mp4} -- $\phi=0.56$ and $\ell_0= 1.0$
    \vspace{-0.2cm}    
    \item \texttt{video\_6.mp4} -- $\phi=0.56$ and $\ell_0= 10.0$
    \vspace{-0.2cm}
    \item \texttt{video\_7.mp4} -- $\phi=0.84$ and $\ell_0= 0.1$
    \vspace{-0.2cm}    
    \item \texttt{video\_8.mp4} -- $\phi=0.84$ and $\ell_0= 1.0$
    \vspace{-0.2cm}
    \item \texttt{video\_9.mp4} -- $\phi=0.84$ and $\ell_0= 10.0$

\end{enumerate}

%\bibliography{SM}% Produces the bibliography via BibTeX.
